\documentclass[12pt,a4paper]{article}
\usepackage{graphics}
\usepackage{amsfonts}

\newcommand{\be}{\begin{equation}}
\newcommand{\ee}{\end{equation}}
\newcommand{\eel}[1]{\label{#1}\end{equation}}
\newcommand{\bea}{\begin{eqnarray}}
\newcommand{\eea}{\end{eqnarray}}
\newcommand{\eeal}[1]{\label{#1}\end{eqnarray}}
\newcommand{\baq}{\begin{equation}\begin{array}{rcl}}
\newcommand{\eaq}{\end{aryray}\end{equation}}
\newcommand{\eaql}[1]{\end{array}\label{#1}\end{equation}}
\newcommand{\beac}{\begin{equation}\begin{array}{rcl}}
\newcommand{\eeacn}[1]{\end{array}\label{#1}\end{equation}}
\newcommand{\ba}{\begin{array}}
\newcommand{\ea}{\end{array}}
\newcommand{\non}{\nonumber \\}
\newcommand{\equ}[1]{(\ref{#1})}

\renewcommand{\a}{\alpha}

\newcommand{\e}{\varepsilon}

\newcommand{\journal}[4]{{\rm #1~}{\bf #2}\,(#3)\,#4}

\newcommand{\pl}{\journal {Phys. Lett.}}

\newcommand{\preprint}[1]{\begin{table}[t]  
           \begin{flushright}               
           \begin{large}{#1}\end{large}     
           \end{flushright}                 
           \end{table}} 

\begin{document}

\begin{titlepage}

\preprint{hep-th/9803137\\TAUP-2499-98}

\vspace{2cm}

\begin{center}
{\bf\LARGE Wilson Loops in the Large $N$ Limit\\
 at Finite Temperature}\footnote{
Work supported in part by the US-Israel Binational Science
 Foundation,
by GIF - the German-Israeli Foundation for Scientific Research,
 and by
the Israel Science Foundation.
}\\
\vspace{1.5cm}
{\bf A. Brandhuber, N. Itzhaki, }\\
{\bf J. Sonnenschein and S. Yankielowicz}\\
\vspace{.7cm}
{\em Raymond and Beverly Sackler Faculty of Exact Sciences\\
 School of Physics and Astronomy\\
 Tel Aviv University, Ramat Aviv, 69978, Israel\\
e-mail: andreasb, sanny, cobi, shimonya@post.tau.ac.il}
\end{center}
\vskip 0.61 cm

\begin{abstract}
Using a proposal of Maldacena we compute in the framework of the
supergravity description of $N$ coincident $D3$ branes the 
energy of a quark anti-quark pair in the large $N$ limit 
of $U(N)$ ${\cal N}=4$ SYM in four dimensions at finite 
temperature. 
\end{abstract}
 
\end{titlepage}

\baselineskip 18pt

Recently, Maldacena conjectured \cite{mal} that the large $N$
limit of certain super-conformal theories is dual to M/string theory 
on a background of AdS times a sphere. 
This remarkable conjecture was studied further in a large number of
papers in the last couple of months with promising results
\cite{hyun}-\cite{gomis}.
In particular, a way how to compute the Wilson line in four
dimensional SYM via supergravity was suggested in \cite{juan1} 
and \cite{sjr}.

In the present work we want to study the finite temperature effects 
on the Wilson line.
We will concentrate on ${\cal N}=4$ SYM in four dimensions.
The difference between the zero temperature case treated in \cite{juan1,sjr}
and the finite temperature case is that now the relevant solution is the near
extremal solution which has the following form
\footnote{
By Wick rotation the metric is transformed into the Euclidean 
solution which has a periodicity along the time direction. The
periodicity is simply the inverse of the Hawking temperature of the
near extremal solution. As a result the two point correlation
function for scalars computed in \cite{gkp,wit} is easily generalized
to the finite temperature case by super-position of ``free" $T = 0$
propagators. The superposition is chosen to ensure the periodicity.}
\cite{mal}
\bea
&& ds^2  =  \alpha' \left\{ \frac{U^2}{R^2} 
[ - f(U) dt^2 + dx_i^2] + R^2 f(U)^{-1}
    \frac{dU^2}{U^2} + R^2 d\Omega_5^2 \right\} ,\non
&& f(U)  =  1 - U^4_T/U^4, \\
&& R^2 = \sqrt{4 \pi g N} ,~~~~~~~~~~~ U_T^4 = 
\frac{2^7}{3} \pi^4 g^2 \mu ~,\nonumber
\eeal{metric}
where $\mu$ is the energy density above extremality on the brane and
the Hawking temperature derived from the Euclidean metric is $T =
U_T/(\pi R^2)$.
Thus what one has to do is simply to follow the line of arguments
in \cite{juan1,sjr}
but with the metric \equ{metric} as a starting point.
However, as we shall see, new ingredients will appear.
Note that for large $R$ at the region outside the horizon the curvature
in string units is small and hence one can trust the supergravity solution
in that region \cite{horo}. At $U=0$ there is a curvature singularity.
However, the region inside the horizon plays no role here.

The action for the string worldsheet is just the usual Nambu-Goto
action
\be 
S = \frac{1}{2 \pi \alpha'} \int d\tau d\sigma \sqrt{h},
\ee
where $h$ is the induced metric on the string worldsheet. Using the
Euclidean form of the metric \equ{metric} as the background we
obtain (in static gauge) the following action
\be
S = \frac{T}{2 \pi} \int dx \sqrt{(\partial_x U)^2 + (U^4 -
U_T^4)/R^4} ~~.
\eel{action}
The action does not depend on $x$ explicitly thus the Hamiltonian in 
the $x$ direction is a constant of motion.
Namely,
\be
\frac{U^4 - U_T^4}{\sqrt{(\partial_x U)^2 + (U^4 - U_T^4)/R^4}} = 
\mbox{const.}= R^2
\sqrt{U_0^4 - U_T^4},
\ee
where the integration constant $U_0$ is the minimal value of $U$ 
which occurs at $x = 0$. This allows us to express $x$ as a
function of $U$ 
\be
x = \frac{R^2}{U_0^3} \sqrt{U_0^4 - U_T^4}
 \int_1^{U/U_0}
\frac{dy}{\sqrt{(y^4 - 1)(y^4 - U_T^4/U_0^4)}} ~.
\ee
The integration constant $U_0$ can, therefore, be related
 to $L$,
the distance between the quark and the anti-quark.
\be
L  =  2\frac{R^2}{U_0} \sqrt{\e} \int_1^{\infty}
\frac{dy}{\sqrt{(y^4 - 1)(y^4 - 1 + \e)}} 
\eel{length}
where $\e = 1 - U_T^4/U_0^4$.

The calculation of the energy proceeds as explained in
\cite{juan1}. To obtain a finite result from \equ{action}
we have to subtract the (infinite) mass of the W-boson which 
corresponds to a string stretched between the brane at 
$U = \infty$ and the $N$ branes.
In the presence of a finite energy the string ends at the 
horizon, $U = U_T$, and not at  $U = 0$.
As we shall see this point is crucial for our discussion.
There are several arguments for this.
The first argument is due to D-branes probing black holes. 
In \cite{jm} it was shown that in the case of finite temperature
the coordinates of the supergravity solution are not identical to
the coordinates of the field theory living on the D-brane.
A coordinate transformation is needed to match them.
This transformation is such that from the point of view of
the field theory living on the brane (at the one-loop order)
the horizon is the origin.
Another argument is that  the Euclidean solution (which is obtained by
wick rotation of \equ{metric}) contains only the region outside the horizon. 
Our last argument is due to Hawking radiation.
As is well known, due to the red shift-effect, the local
 temperature close to the horizon is very high.
In fact it is so high that any static particle/string will burn.
In our case, by comparing the local temperature, $T_{loc}\sim
T_{Haw}\sqrt{g^{tt}}$ to the string mass, $M_s=1/\sqrt{\a'}$ 
we find that the minimal distance for the string not to
burn is $U_T(1+1/R^2)$.
Since the supergravity description is valid for large $R$ we 
conclude that the ends are at $U=U_T$.
Integrating from the horizon we obtain a finite result for the 
static energy 
\be
E = \frac{1}{\pi} \left\{ U_0 \int_1^\infty \left(\frac{\sqrt{y^4 - 1 +
      \e}}{\sqrt{y^4 - 1}} - 1 \right) - U_0 + U_T \right\}. 
\eel{energy}

\begin{figure}[h!]
\begin{tabular}{p{0.45\textwidth}p{0.45\textwidth}}
 \resizebox{0.45\textwidth}{!}{\includegraphics{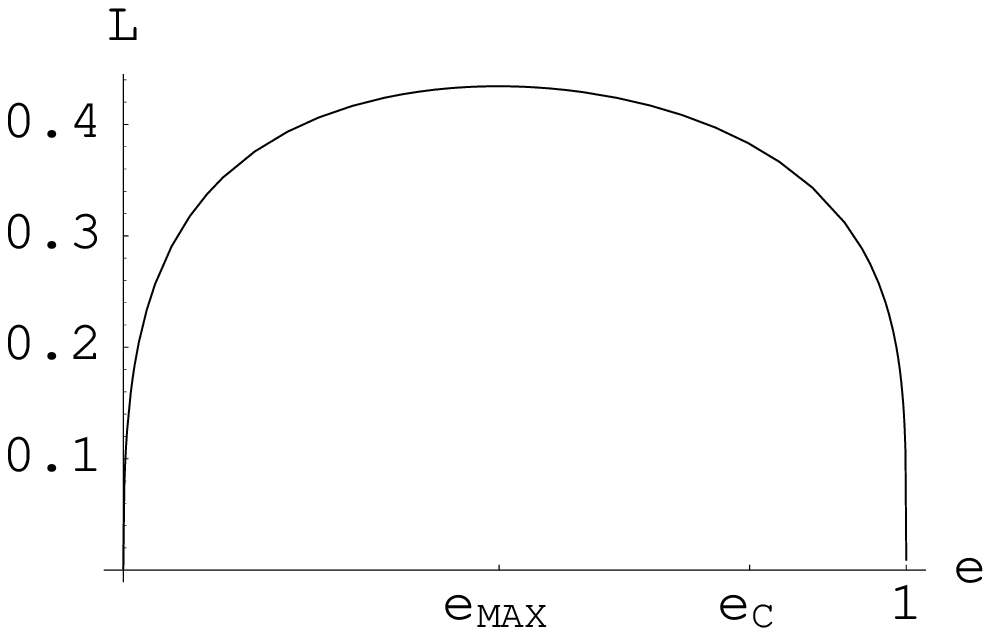}}
&
 \resizebox{0.45\textwidth}{!}{\includegraphics{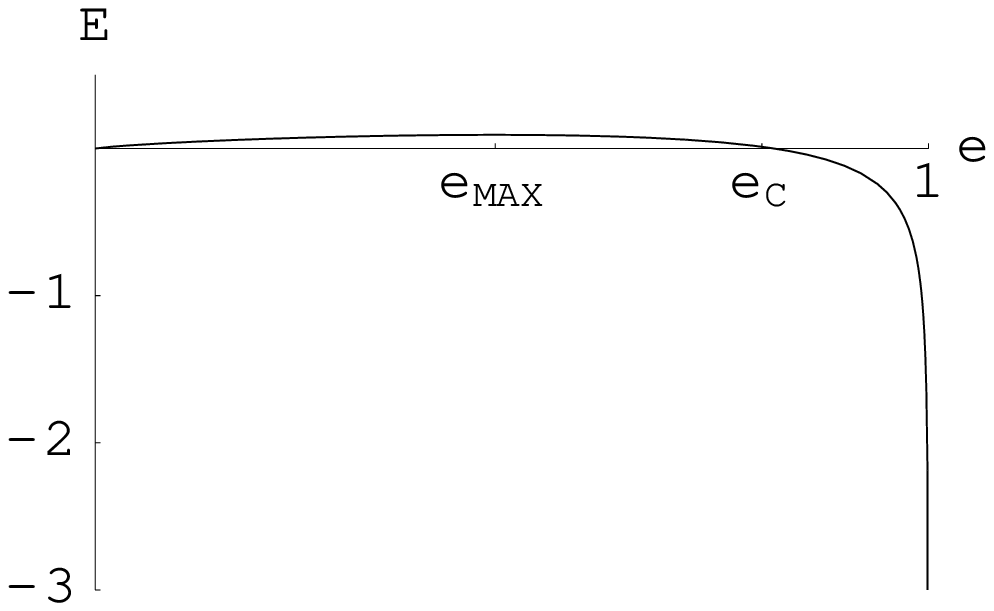}}
\\
\caption{The distance $L$ between the quark-anti-quark pair as a
  function of $\epsilon = \epsilon(T,\tilde U)$. Note that $L$ has a
  maximal value $L_{max}$.}
&
\caption{The energy of the quark-anti-quark pair relative to the
  ``free" quark situation  as a function of $\epsilon =
  \epsilon(T,\tilde U)$. Note that $E = 0$ is achieved at $\epsilon_C
  < \epsilon_{max}$ corresponding to $L_C < L_{max}$.}
\end{tabular}
\end{figure}
What we are after is the static energy $E(L,T)$ between the
 ``quark'' and the ``anti-quark''.
To obtain this expression we have to eliminate $U_0$ between 
eqs. \equ{energy} and \equ{length}.
This can be done only numerically.
Instead we shall find the qualitative behavior by looking at 
\equ{energy} and \equ{length} and the corresponding numerical
integration depicted in fig. 1 and fig. 2.
We note that $\epsilon\approx 1$ corresponds to the low temperature
region, $TL\ll 1$, while  $\epsilon\approx 0$ corresponds to the
high temperature region $TL \gg 1$.

For small temperature the behavior is roughly $E \sim -1/L$ as
in the
zero temperature case \cite{juan1}. Taking into account the lowest
corrections in $U_0$ the expression is
\be
E = -{2\sqrt{2}\pi^{3/2}(4\pi g^2_{YM}N)^{1/2}\over
\Gamma(1/4)^4}{1\over L}[1 + c(TL)^4] 
\ee
where $c$ is a positive numerical constant which does not depend on $R$.
The underlying conformal nature of the theory reveals itself in the 
fact that $E L$ can depend on $T$ only through the combination $T L$.

The behavior of $L$ as a function of $\epsilon$ seems a priori
puzzling since it indicates the existence of a maximum distance
$L_{max}$. Indeed, if we assume that the behavior depicted in fig. 1
and fig. 2 always holds we will run into strange double valued
behavior of $E(L,T)$ for $L > L_{max}$. Fortunately physics tells us
to believe the result only in the region $0<L<L_C$ where $L_C < L_{max}$.
The existence of $L_C$ is seen in fig. 2. Starting from the low
temperature region $\epsilon \sim 1$ we reach $\epsilon_C$ at 
which $E = 0$. At this point the energy associated with our string
configuration (fig. 3) is the same as the energy of a pair of 
free quark and anti-quark with asymptotically zero force
between them (fig. 4). It is important to note that $\epsilon_C$ is
reached before $L$ reaches its maximal value $L_{max}$ (fig. 1).
Once we reach $L_C$ our string configuration (fig. 3) does not
correspond to the lowest energy configuration and we should stop
to trust eqs. \equ{length} and \equ{energy}.

The physical picture which emerges is quite reasonable and
simple. For a given temperature $T$ we encounter two regions with
different behavior. For $L << 1/T$ we observe a Coulomb like
behavior while for $L >> 1/T$ the quarks become free due to 
screening by the thermal bath.

\begin{figure}[h!]
\begin{center}
\begin{tabular}{p{0.3\textwidth}p{0.3\textwidth}}
 \resizebox{0.3\textwidth}{!}{\includegraphics{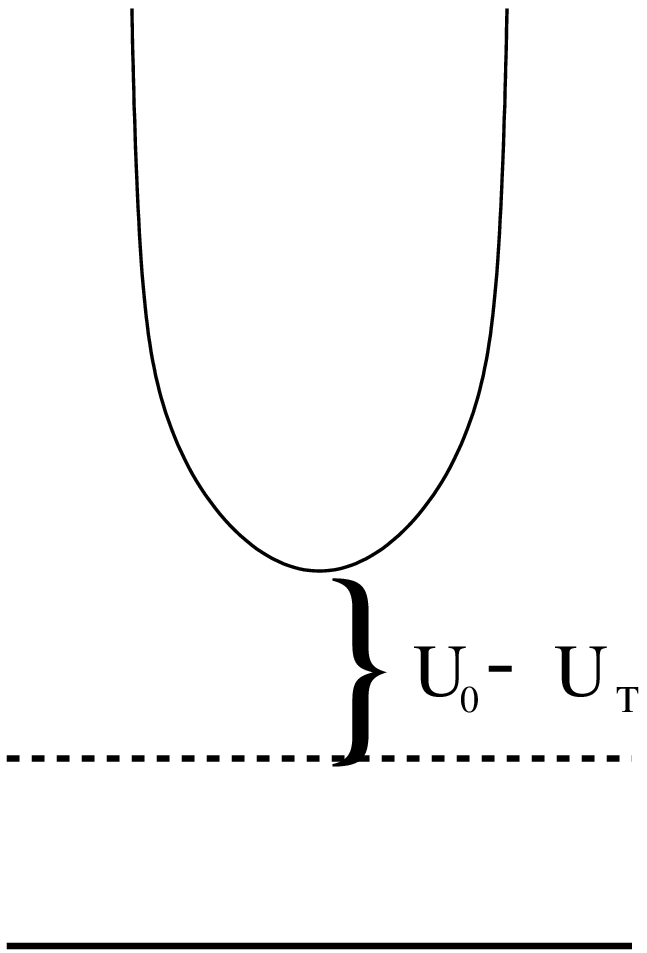}}
&
 \resizebox{0.3\textwidth}{!}{\includegraphics{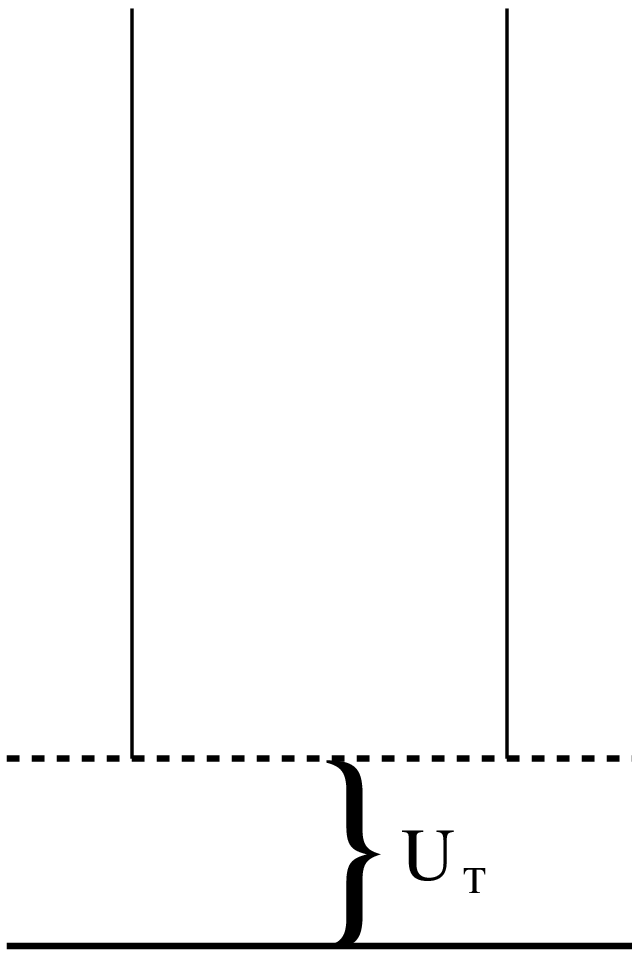}}
\\
\caption{The energetically favorable configuration for $L < L_C$.}
&
\caption{The energetically favorable configuration for $L > L_C$.}
\end{tabular}
\end{center}
\end{figure}
In fig 5. we have plotted $E = E(L)$ for a given $T$ by eliminating
$\epsilon$ between eqs. \equ{length} and \equ{energy} and trusting the
result up to $L_C$.
\begin{figure}[h!]
\begin{center}
\resizebox{0.4\textwidth}{!}{\includegraphics{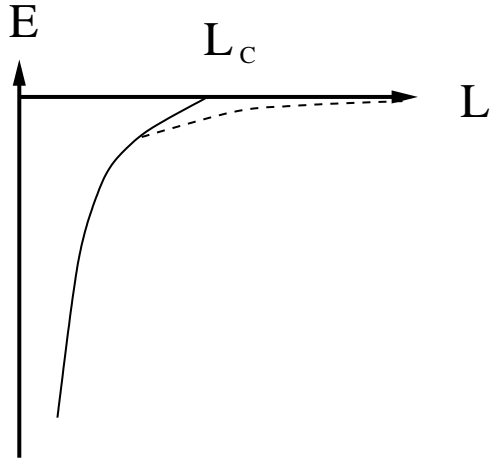}}
\end{center}
\caption{The energy $E$ of the quark-anti-quark pair as a function of
  $L$ for a given $T$. The solid line corresponds to the numerical
  calculation up to $L_C$, the dashed line indicates the expected
  behavior for large $L$.}
\end{figure}

The non-conformal theories, studied in
\cite{juan} from the supergravity point of view, 
contain a length scale (which is related to $g_{YM}$) and as such 
a phase transition might take place.
It should be interesting to study these phase transitions and their relation
to the transitions between the supergravity description and the 
perturbative/conformal field theory description discussed in \cite{juan}.
 
Our result agrees with observations made in \cite{witten1} about the
conformal theories, namely, that there are no phase transitions at finite
temperature. This is only true if the theory lives on a non-compact
space which is the case in our paper. In \cite{witten1} it was shown that the
same theory on $S^3$ shows a confinement/deconfinement phase
transition because the radius of the sphere introduces a scale and
breaks conformal invariance. 

\begin{center}{\bf Acknowledgements}\end{center}

It is a pleasure to thank J. Maldacena for helpful correspondence. 
While working on this paper we were informed by
S.J. Rey, S. Theisen and J.T. Yee, that they are studying very similar 
questions in a forthcoming work. Very recently a related preprint
by E. Witten (\textrm{hep-th/9803131}) appeared. 


\end{document}